\documentclass[prl,twocolumn,superscriptaddress,floatfix]{revtex4}
\usepackage{comment}
\usepackage{hyperref}
\usepackage{lipsum}
\usepackage[british]{babel}
\usepackage[utf8]{inputenc}
\usepackage{amsmath}
\usepackage{amssymb}
\usepackage{mathtools}
\usepackage{braket}
\usepackage{graphicx}
\usepackage{xcolor}
\usepackage{braket}
\usepackage[normalem]{ulem}

\begin{document}

\title{Quasinormal Modes of Optical Solitons}
\author{Christopher Burgess}
\affiliation{School of Physics and Astronomy, SUPA, University of St. Andrews,
North Haugh, St. Andrews, KY16 9SS, UK}
\author{Sam Patrick}
\author{Theo Torres}
\affiliation{Department of Physics, King’s College London,
The Strand, London, WC2R 2LS, UK}
\author{Ruth Gregory}
\affiliation{Department of Physics, King’s College London,
The Strand, London, WC2R 2LS, UK}
\affiliation{Perimeter Institute, 31 Caroline Street North, Waterloo, 
ON, N2L 2Y5, Canada}
\author{Friedrich K\"{o}nig}
\email{fewk@st-andrews.ac.uk}
\affiliation{School of Physics and Astronomy, SUPA, University of St. Andrews,
North Haugh, St. Andrews, KY16 9SS, UK}
\date{\today}

\begin{abstract}
Quasinormal modes (QNMs) are essential for understanding the stability and resonances of open systems, with increasing prominence in black hole physics.
We present here the first study of QNMs of optical potentials.
We show that solitons can support QNMs, deriving a soliton perturbation equation and giving exact analytical expressions for the QNMs of fiber solitons.
We discuss the boundary conditions in this intrinsically dispersive system and identify novel signatures of dispersion.
From here, we discover a new analogy with black holes and describe a regime in which the soliton is a robust black hole simulator for light-ring phenomena.
Our results invite a range of applications, from the description of optical pulse propagation with QNMs to the use of state-of-the-art technology from fiber optics to address questions in black hole physics, such as QNM spectral instabilities and the role of nonlinearities in ringdown.
\end{abstract}

\maketitle\noindent

\emph{Introduction}.---Quasinormal modes (QNMs) are an area of high activity and interest following the discovery of gravitational radiation from black hole mergers \cite{LIGO}.
After a merger, the resulting black hole \textit{rings down} to its final state in a characteristic fashion, described by a damped waveform with complex frequency.
This phenomenon of ringdown is a generic feature of open systems, allowing us to identify natural resonances and address questions of stability~\cite{dorey_resonance, konoplya_stability, cardoso_stability}.
Examples of ringdown abound, in optical cavities ~\cite{lalanne,ching,alpeggiani}, plasmonic nanoresonators~\cite{kristensen, yan}, polariton superfluids~\cite{jacquet_polaritons}, hydrodynamics~\cite{torres_qnm,vieira}, supergravity~\cite{kumar_supergravity}, and even the ringing of church bells.
This diversity of settings is crucial as it provides many perspectives on the ringdown phenomenon, whose significance in the context of black holes is increasingly recognised~\cite{cardoso_lypanov,dreyer_immirzi,gossan_nohair}.

The ringdown of a perturbed open system is readily understood using its QNMs.
These are eigenmodes of the evolution operator, with a discrete complex spectrum, $\Omega_n = \omega_n - i\Gamma_n$, where the overtone index $n$ orders the modes by increasing decay rate $\Gamma_n$.
The signal of ringdown is a superposition of damped QNM oscillations, with the ringdown spectrum a feature of the system, independent of the initial perturbation~\cite{kokkotas}.
The fundamental mode ($n = 0$) is longest-lived, eventually dominating the linear response and providing immediate access to characteristic information~\cite{berti}.
QNMs describe both perturbative field evolution around fixed bulk media with well-defined boundaries, and also open systems with time-independent repulsive potentials.
In optics, they efficiently reconstruct the mode shapes of electromagnetic fields in optical resonators and plasmonic cavities, both of the former kind~\cite{lalanne}.
By contrast, black hole oscillations are of the latter type~\cite{torres_qnm,berti}.
To date, QNMs of optical potentials have not been reported.
Fiber optical solitons provide a remarkable way of creating such potentials.
Indeed, suitable perturbations to the soliton are known to obey a Schr\"{o}dinger equation with a repulsive potential in the comoving frame~\cite{gorbach, philbin, amol, wang}.

In this Letter, we begin by deriving a perturbation equation where the soliton acts as a potential, due to a nonlinear polarization of the medium, with weak third-order dispersion at the soliton.
We clarify the notion of QNMs in this intrinsically dispersive system, finding signatures of dispersion, and showing the soliton can support a discrete set.
Analytic expressions for the QNMs and their complex frequencies---the QNM spectrum---are provided.
We then consider a weak dispersive pulse co-propagating with and perturbing the soliton.
Simulating the response, we observe the predicted ringdown. We identify the complex frequency of the fundamental mode, finding agreement with our theory.
Finally, we establish a mathematical analogy between the ringdown of solitons and black holes, and discuss prospects for developing our analysis in this and other soliton-supporting systems.

\emph{Soliton perturbations}.---To investigate dispersive pulse interactions with a soliton, we first describe the soliton in single-mode fibers.
The soliton is given by a complex envelope $A = A_s(z, t)$ around a carrier of frequency $\omega_s$, obeying the non-linear Schr\"{o}dinger equation (NLS) for pulse propagation~\cite{agrawal}.
The soliton has a stationary sech$^2(\tau)$ intensity profile in the comoving frame, which is related to the laboratory frame by
\begin{equation}
\tau = \frac{t - z/v}{T_0}, \quad \zeta = \frac{ z |\beta_{s2}|} {T^2_0}\label{scales},
\end{equation}
with $T_0$ the temporal width of the soliton, $v$ its group velocity, and $\beta_{s2}$ its group velocity dispersion.

In the absence of higher-order dispersion, perturbed solitons are known to relax by routes other than ringdown~\cite{yang,akhmediev_cherenkov}.
Therefore, we consider the NLS with additional terms for higher-order dispersion~\cite{skryabin}.
In the laboratory frame, this reads
\begin{equation}
\frac{\partial A}{\partial z} - i\big[\beta(i\partial_t + \omega_s) - \beta(\omega_s)\big] A - i\gamma|A|^2 A = 0\label{GNLS},
\end{equation}
where $\beta(\omega)$ is a Taylor series for the propagation constant describing dispersion in the fiber, and $\gamma$ is the fiber nonlinear parameter.
We extend the analysis in \cite{amol, wang}, which limited dispersion to second-order around the soliton frequency, by including weak third-order dispersion. This preserves stable soliton propagation, with only the soliton velocity $v$ and phase affected.
The perturbed soliton solution is given in~\cite{hasegawa, akhmediev}.

We consider a dispersive pulse, $a$, as a linear perturbation co-propagating with the soliton, $A_s$.
The dispersive pulse envelope is defined with a carrier frequency, $\omega_a$, so the overall envelope is
\begin{equation}
A = A_s + a\, e^{i(\beta(\omega_a)-\beta(\omega_s))z - i(\omega_a-\omega_s) t}.
\end{equation}
Inserting into Eq.~(\ref{GNLS}) yields the linearized equation of motion for the dispersive pulse.
We neglect fast-oscillating terms arising from frequency mixing between the soliton and dispersive pulse, given their spectral separation and no phase matching.
In the comoving frame, the dispersive pulse envelope satisfies
\begin{align}
&\frac{|\beta_{s2}|}{T_0^2} \frac{\partial a}{\partial \zeta} - \left(\frac{\beta_{s1}}{T_0} -\frac{\beta_{s3}}{6 T^3_0} - \frac{\beta_{a1}}{T_0}\right)\frac{\partial a}{\partial \tau} \nonumber \\&\quad\quad\quad - i  \sum_{k=2}^{\infty}\frac{\beta_{ak}}{k!}\left(\frac{i\partial_{\tau}}{T_0}\right)^{\!\!k} a - 2i\gamma |A_s|^2 a = 0,\label{PE1}
\end{align}
with $\beta_{sk}$ and $\beta_{ak}$ the expansion coefficients of $\beta(\omega)$ about $\omega_s$ and $\omega_a$, respectively.
Requiring the dispersive pulse carrier to be group-velocity matched (GVM) to the soliton, so that $\beta_{a1} = \beta_{s1} -\beta_{s3}/6 T^2_0 = v^{-1}$, the equation simplifies.
For narrowband perturbations, we again truncate our Taylor series for dispersion, but now to second-order around $\omega_a$.
We thus focus on near-GVM perturbations. This is natural in the context of QNMs, which are associated with perturbations that remain coincident with a background potential until late times.
These perturbations satisfy
\begin{align}
i\frac{\partial a}{\partial \zeta} - \frac{\beta_{a2}}{2  |\beta_{s2}|}\frac{\partial^2 a}{\partial \tau^2} + 2 \ \! \textnormal{sech}^2(\tau) a = 0,\label{PE2}
\end{align}
having inserted the soliton~\cite{hasegawa, akhmediev}.
Eq.~(\ref{PE2}) is our soliton perturbation equation, which may be cast as a time-reversed Schr\"{o}dinger equation with an inverted P\"{o}schl-Teller potential~\cite{poschl}.
This agrees with previous work in which the soliton acted as a repulsive potential, creating classical turning points for slow dispersive light~\cite{amol, tartara}, and exhibiting light tunneling~\cite{philbin}.

\begin{figure}
\centering
\includegraphics[width=245pt]{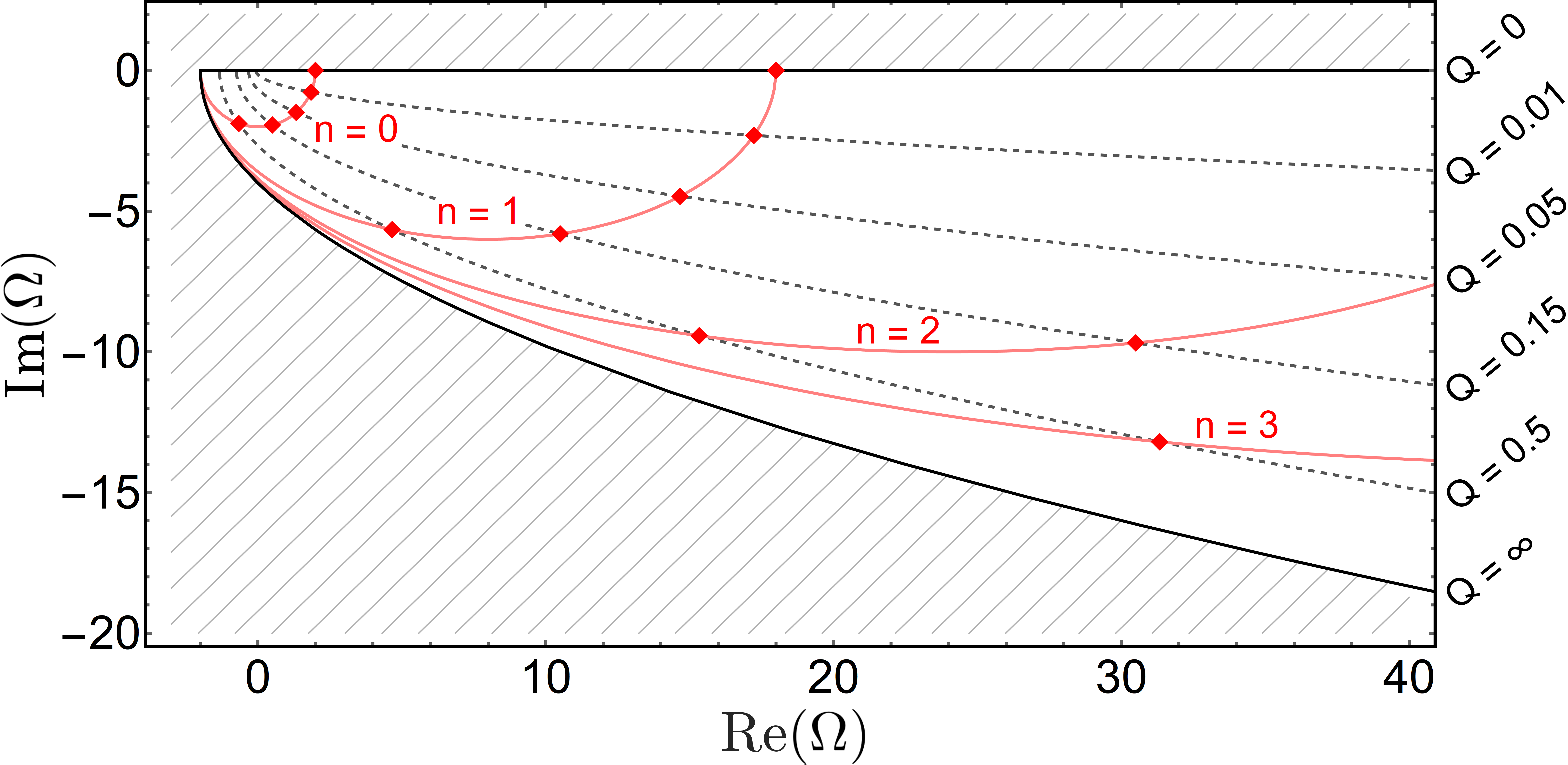}
\caption{
Quasinormal mode spectra for optical solitons with $Q = 4|\beta_{s2}|/\beta_{a2} - 1/4$ in the range $[0,\infty)$, plotted in the lower-half complex plane.
Each QNM frequency ({\color{red}red} diamond) lies at the intersection of a parabola ({\color{black!65!white}grey}, dashed) corresponding to $Q$, and a semi-ellipse ({\color{red}red}, solid) corresponding to the overtone index $n$.
The fundamental mode frequencies lie on a semi-circle of radius 2 about the origin.
}
\label{fig:FPmodes1}
\end{figure}

\begin{figure*}[t]
\centering
\includegraphics[width=\textwidth]{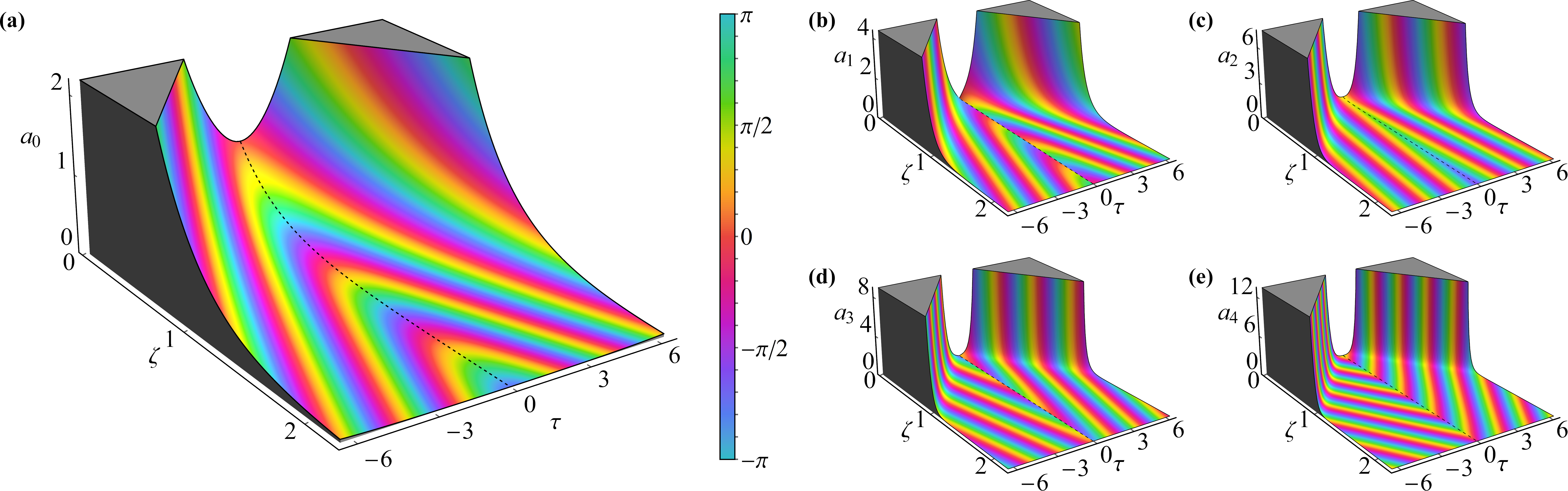}
\caption{
Complex plots of soliton quasinormal modes, $a_n(\zeta, \tau)$ in units of $\alpha$.
The fundamental mode (\textbf{a}) and four overtones \textbf{(b-e)} are shown for $\zeta \in [0,2], \tau \in [-6,6]$ with $|\beta_{s2}|/\beta_{a2} = 2$.
They grow symmetrically about the soliton potential at $\tau = 0$, and their phase velocities reverse direction for sufficiently high overtones, visible from the contours of constant phase.
This signature of dispersion represents a qualitative departure from the quasinormal modes of non-dispersive systems.}
\label{fig:FPmodes2}
\end{figure*}

\emph{Soliton quasinormal modes}.---As ringdown arises in the linear response of a system, soliton ringdown can be explored using our soliton perturbation equation.
To derive the soliton QNMs, we begin with mode solutions, $a(\zeta, \tau) = u(\tau)\exp(-i\Omega\zeta)$.
We obtain
\begin{align}
- \frac{d^2 u}{d \tau^2} + \frac{2 |\beta_{s2}| }{\beta_{a2}} \bigg[ \Omega+ 2 \ \! \textnormal{sech}^2(\tau)\bigg] u = 0,\label{QNMeqn}
\end{align}
solvable through exact methods~\cite{landau}.

Next, we must impose boundary conditions on the general solution to Eq.~\eqref{QNMeqn}.
In gravitational physics, one requires outgoing waves at the boundaries, as phase and group velocities are equivalent in relativistic systems, guaranteeing energy-dissipating modes.
However, the situation is not so simple in the presence of dispersion.
Moreover, laboratory frame energy is not conserved by Eq.~\eqref{PE2}, as time translation is not a symmetry.
The relevant symmetry is translation in $\zeta$, generating the conserved current,
\begin{equation}
    j = -\frac{\beta_{a2}}{2  |\beta_{s2}| }\bigg[\frac{\partial a}{\partial \zeta}\frac{\partial a^*}{\partial\tau} + \frac{\partial a^*}{\partial \zeta}\frac{\partial a}{\partial\tau}\bigg].\label{current}
\end{equation}
It is then natural that we require $j$ to be asymptotically outgoing, i.e. $\text{sgn}(j) = \text{sgn}(\tau)$ as $\tau\to\pm\infty$.
In the absence of dispersion, these boundary conditions coincide with those of relativistic systems.
Far from the soliton potential, $a(\zeta, \tau) \sim \exp(iK_{\pm}\tau - i\Omega\zeta)$, with asymptotic wavenumbers $K_{\pm}$ given by the fiber's underlying dispersion relation, $\Omega = -\frac{\beta_{a2}}{2|\beta_{s2}|}K_{\pm}^2$, due to Eq.~(\ref{PE2}).
For decaying modes, i.e. $\text{Im}(\Omega) < 0$, the boundary conditions thus set the signs of $\text{Im}(K_+)$ and $\text{Im}(K_-)$ so that solutions grow exponentially towards the boundaries.

For complex $\Omega$, the asymptotic form of the general mode solution contains waves that both grow and decay with separation from the soliton.
QNM boundary conditions forbid the latter, as they deliver energy into the system.
These unwanted waves vanish only for a discrete set of QNM frequencies $\Omega_n$ where the Wronskian of two linearly independent solutions is zero, and decaying waves vanish against growing waves.
In the language of scattering theory, this occurs due to the divergence of transmission and reflection amplitudes at these frequencies.
These frequencies also appear as poles in the Green functions associated with Eq.~(\ref{PE2}).
Therefore, our physically motivated condition that $j$ be outgoing at infinity agrees with the standard definition of QNMs, reinforcing our approach to the boundary conditions~\cite{nollert2, leaver, andersson}.

The QNM frequencies are given by
\begin{align}
\Omega_n &= \frac{\beta_{a2}}{2  |\beta_{s2}|}\Bigg[\bigg(n+\frac{1}{2}\bigg)^{\!\! 2} - \bigg(\frac{4|\beta_{s2}|}{\beta_{a2}}-\frac{1}{4}\bigg)\Bigg]\nonumber\\
&\quad-i\frac{\beta_{a2}}{  |\beta_{s2}|}\bigg(n+\frac{1}{2}\bigg)\sqrt{\frac{4|\beta_{s2}|}{\beta_{a2}}-\frac{1}{4}} \label{QNMfreq},
\end{align}
where the overtone index $n$ is a non-negative integer. The soliton QNM spectrum, plotted in Fig.~\ref{fig:FPmodes1}, depends only on the group velocity dispersions, not on $\beta_{s3}$.
This is tunable by varying the soliton central frequency, which determines the dispersion at the soliton and GVM frequencies.
From Eq.~(\ref{QNMfreq}), we see the soliton supports QNMs only under normal dispersion, $\beta_{a2}>0$, confirming that the perturbation must be spectrally distinct from the anomalous dispersion regime of the soliton, $\beta_{s2}<0$.

The full QNM solution is easily obtained with the general solution and QNM frequencies. The result is
\begin{align}
a_n(\zeta, \tau) &= \alpha \cosh^{n+\frac{1}{2}}(\tau) e^{\text{Im}(\Omega_n)\zeta}  e^{i\phi_n(\tau,\zeta)} f_n(\tau),\label{QNMSolution}
\end{align}
where $\alpha$ is the amplitude and we define
\begin{align}
\phi_n(\tau,\zeta) &\equiv -\text{Im}(\lambda)\log \cosh(\tau) - \text{Re}(\Omega_n)\zeta, \nonumber\\
f_n(\tau) &\equiv\!\!\ \,_2F_1\big[{-}n, 1+2\lambda-n;1+\lambda-n;(e^{2\tau} + 1)^{-1}\big],\nonumber\\
\lambda &\equiv -\frac{1}{2} + i\sqrt{\frac{4|\beta_{s2}|}{\beta_{a2}} - \frac{1}{4}}.\nonumber
\end{align}
The hyperbolic factor in Eq.~(\ref{QNMSolution}) shapes the mode with exponential growth far from the soliton, while the exponential in $\zeta$ gives an overall decay.
The phase $\phi_n$ produces phase velocities that are outgoing for the lowest overtones, but otherwise are ingoing: a result due to dispersion. The ordinary hypergeometric function $f_n$ arranges that mode parity alternates with overtone index. For the fundamental QNM, $f_0 = 1$.
The first five QNMs are plotted in Fig.~\ref{fig:FPmodes2}.
The exponential growth of these solutions appears unphysical, but this is a typical feature of QNMs, which are necessarily decaying in advanced time. Physically, a field ringing down resembles a superposition of QNMs on only a finite region of the space~\cite{kokkotas}.

\emph{Simulations}.---To demonstrate the emission of ringdown waves from the soliton, we numerically simulate the evolution of a near-GVM pulse colliding with the soliton.
For various initial pulses, the collision produces ringdown waves in the perturbative pulse field.
These waves visit each position in the comoving frame, in both transmission and reflection. 
At each position, the signature of ringdown is a decaying oscillation, with a period and decay rate given by the fundamental QNM.
We simulate this process using a split-step Fourier method to solve the soliton perturbation equation, Eq.~(\ref{PE2}).

The ringdown of an optical soliton occurs against an evolving background of non-QNM contributions to the perturbative field.
These arise due to dispersion, which tends to broaden the transmitted and reflected pulses produced by the pulse-soliton collision.
This effect is absent in the paradigmatic QNM systems~\cite{lalanne,kristensen, yan,jacquet_polaritons,torres_qnm,kumar_supergravity}.
The ringdown signal is strongest around an observation point $\tau = \tau_0$ near the transmitted pulse, and clearest when this pulse has least width.
Therefore, we configure our initial pulse with a quadratic phase, known as a chirp, so the transmitted pulse is narrowest at a point $\tau = \tau_c$ near the observation point. Our initial condition is
\begin{equation}\label{eq:IC}
    a(0, \tau) = \alpha\exp\!\bigg[{-}\frac{(\tau-\tau_p)^{2p}}{\vphantom{(\tau-\tau_p)^{2p}_{2p}}\sigma^{2p}}\bigg]\exp\!\bigg[i\frac{(\tau-\tau_c)^2\vphantom{(\tau-\tau_p)^{2p}}}{4\zeta_c\vphantom{(\tau-\tau_p)^{2p}}}\bigg],
\end{equation}
with a super-Gaussian pulse envelope about $\tau = \tau_p$ and a minimum pulse width at a fiber length of $\zeta = \zeta_c$.

\begin{figure}
\centering
\includegraphics[width=\textwidth/2]{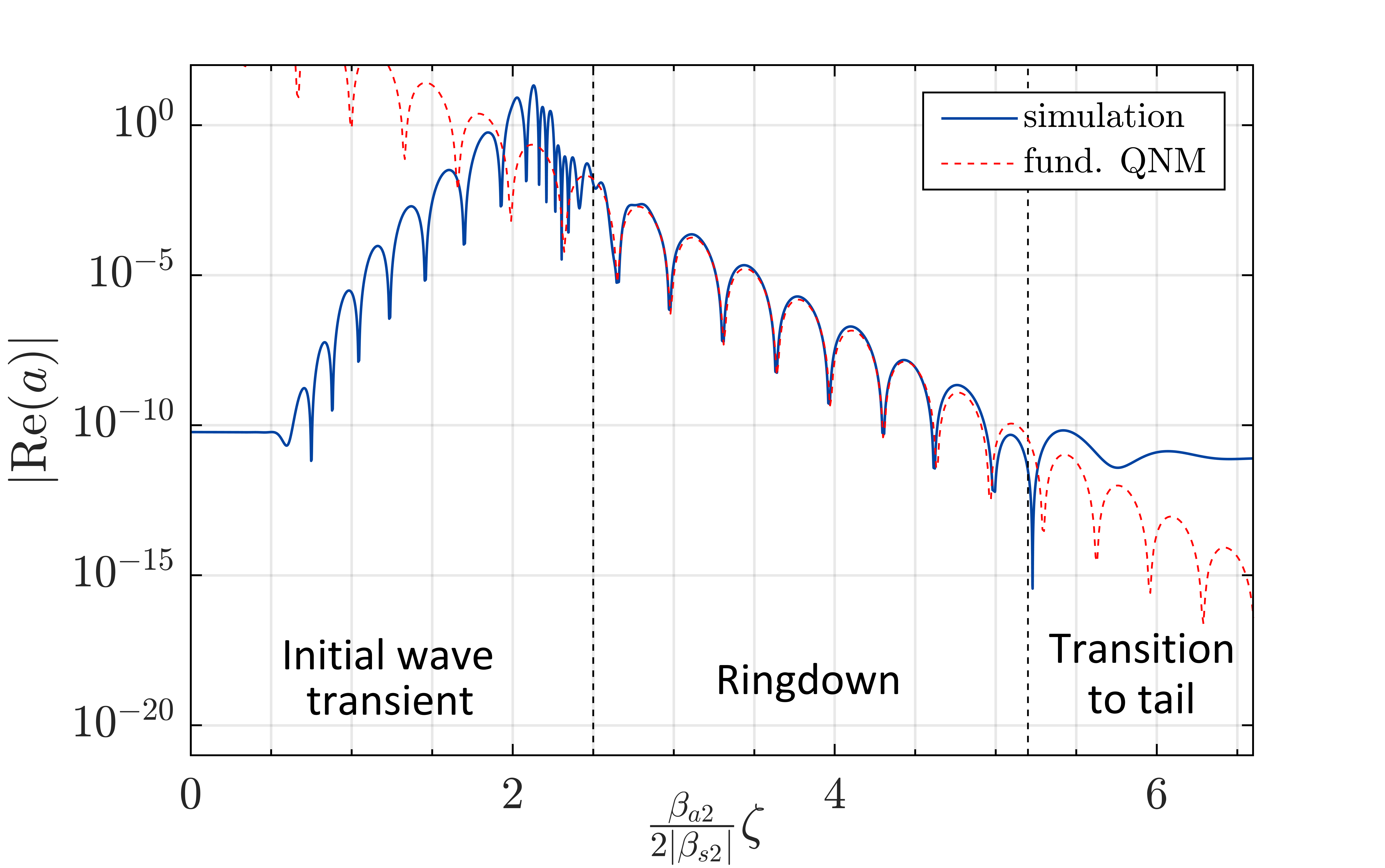}
\caption{
Evolution of the perturbative field $a(\zeta, \tau)$ at a fixed point $\tau_0=-5$ away from the soliton, in units of $\alpha$.
The blue curve is the simulated solution to Eq.~(\ref{PE2}) with the initial condition in Eq.~(\ref{eq:IC}), $\tau_c = -10$, $\tau_p = 20$, $\zeta_c=2.5$, $p=5$, $\sigma = 15$ and $|\beta_{s2}|/\beta_{a2}=2.5$.
Three phases are separated by vertical dashed lines, corresponding to i) the initial transient of the perturbation, ii) the emission of ringdown waves, and iii) a transition to a late-time tail.
The red dashed curve shows a fit with the fundamental QNM ($n=0$) in Eq.~(\ref{QNMfreq}).}
\label{fig:QNM_ringdown}
\end{figure}

The above initial condition produces a clear ringdown signal for a range of parameters.
Importantly, the dominant ringdown period and decay rate consistently agree with the fundamental mode.
Fig.~\ref{fig:QNM_ringdown} presents a quintessential example of optical soliton ringdown.
Three phases are identifiable: i) an initial transient phase as the transmitted pulse passes the observation point, ii) a relaxation phase dominated by ringdown waves, and iii) a transition towards a late-time power law decay owing to dispersion.
These three phases are analogous to those appearing in the relaxation of black holes or hydrodynamic vortex flows~\cite{konoplya_stability,berti,torres_qnm}.
For non-dispersive systems, the presence of a late-time tail depends only on details of the potential.
In contrast, we attribute ours to the action of dispersion, as the inverted P\"{o}schl-Teller potential does not otherwise exhibit a late-time power law decay~\cite{casals}.

\emph{Black hole analogy}.---In optics, ringdown is predominately associated with leaky cavities and resonators.
Yet the ringdown of an optical soliton, viewed as a repulsive potential, bears greater resemblance to that of black holes.
Indeed, the link to black holes is beyond merely qualitative.
Eq.~(\ref{QNMeqn}) is an exact mathematical analogue to the QNM equation for scalar, electromagnetic and gravitational perturbations on the Nariai and near-extremal Schwarzschild-de Sitter black holes~\cite{konoplya_stability},
\begin{align}
- \frac{d^2 R}{d r_*^2} + \Big[{-}\omega^2+ V_0 \ \! \textnormal{sech}^2(\sigma r_*)\Big] R = 0,\label{BHeqn}
\end{align}
with $R$ the radial wavefunction and $r_*$ the so-called tortoise co-ordinate.
Importantly, Eq.~(\ref{BHeqn}) is routinely used to produce analytic approximations for the QNM frequencies of a wide range of black holes, modelling the effective gravitational potential with the P\"{o}schl-Teller potential. Indeed, this approach is common in the astrophysical context, applying to both Schwarzschild and Kerr black holes~\cite{berti}.

In the analogy, the soliton reproduces the entire effective radial gravitational potential outside of the black hole, with the centre of the soliton corresponding to the photon sphere.
The centre of the soliton is a point of instability that gradually leaks near-GVM perturbations, just as the photon sphere leaks light.
The mode shapes are identical to those of the soliton, with frequencies related by $\omega \propto \sqrt{-\Omega}$.
Consequently, whereas the black hole QNM frequencies lie along vertical lines in the complex plane, those of the soliton form parabolic curves, shown in Fig.~\ref{fig:FPmodes1}.
We can define a regime of weak dispersion in which the group velocity dispersion at the GVM frequency is much less than the finite dispersion supporting the soliton, i.e., $\beta_{a2} \ll |\beta_{s2}|$. The low-lying QNM spectrum is then given by
\begin{equation}
\Omega_n = {-}2 - i\sqrt{\frac{4\beta_{a2}}{|\beta_{s2}|}}\bigg(n+\frac{1}{2}\bigg),
\end{equation}
where $n$ is small compared with $\sqrt{|\beta_{s2}|/\beta_{a2}}$ and we omit terms of order $O(\beta_{a2}/|\beta_{s2}|)$.
After re-scaling, this is formally identical to the QNM spectrum of the Nariai and near-extremal Schwarzschild-de Sitter black holes, in the regime of high angular momentum perturbations, i.e., the eikonal regime \cite{konoplya_stability, berti}.
Thus, the soliton ringdown is identical to that of these black holes, and the soliton becomes a black hole simulator.

\emph{Outlook}.---We have shown for the first time that optical potentials can support quasi-normal modes and demonstrated how ringdown may be excited in the case of fiber solitons. We have demonstrated a selection of correspondences, contributing to a growing zoo of table-top systems in analogue gravity~\cite{jacquet_analogues}.
In particular, this work sounds a bell, heralding future works that will introduce sophisticated techniques of QNM analysis from black hole physics into optics.
Conversely, fiber solitons are now a platform with which the tools of optics may now be put to use in the study of otherwise inaccessible phenomena that arise in the study of black holes, such as the recently reported QNM spectral instability~\cite{jaramillo, alsheikh, berti2, torres_instability} and imprints of nonlinearity on the ringdown process \cite{mitman,cheung}.

The above analysis employed several approximations that may be fruitful to relax, to probe the consequences for soliton QNMs.
In this work, the soliton QNM spectrum was shown to be robust against weak third-order dispersion at the central frequency of the soliton.
Future work can investigate the influence of further weak fiber optical effects, e.g.~Raman scattering and higher-order dispersion at the perturbation frequency.

We generalized the QNM boundary conditions to an outgoing energy current, accommodating dispersion in our system.
We discovered QNMs with ingoing phase velocities and dispersion-induced late-time tails, revealing the soliton as a platform for investigating dispersive QNM physics.
A further effect of dispersion is that the soliton QNM spectrum is tunable by varying the central frequency of the soliton.
In fibers supporting several group velocity matched points, there exists a further discrete freedom to choose the GVM frequency around which we consider perturbations.
The same soliton may thus possess several distinct QNM spectra.
Beyond this, we can relax our near-GVM condition and consider QNMs of more general linear differential operators in Eq.~(\ref{PE1}).
This can be expected to produce a continuum of QNM spectra, fully characterising the ringdown processes available to the soliton.

The above QNM analysis focused on optical solitons, but our findings can describe natural resonances in any realistic system supporting NLS-type solitons, providing immediate extensions beyond optics.
Furthermore, this work lays the foundation for determining QNMs of other solitons (e.g.\ KdV, sine-Gordon), wherever they act as stationary effective potentials in their comoving frame.

Finally, we note that optical soliton ringdown has yet to be experimentally observed and tested against the framework of QNMs.
This is most promising with ultra-short pulses, because the relevant length scales grow quadratically with pulse length, and would otherwise extend beyond the fiber loss length.
We anticipate that the identification of ringdown waves in propagating optical pulses will stimulate developments in communications and ultra-fast lasers, and motivate advances in the field of dispersion engineering.

For the purpose of open access, the authors have applied a creative commons attribution (CC BY) licence (where permitted by UKRI, ‘open government licence’ or ‘creative commons attribution no-derivatives (CC BY-ND) licence’ may be stated instead) to any author accepted manuscript version arising. The supporting data for this Letter are openly available from \cite{data_source}.

This work was supported in part by the Science and Technology Facilities Council through the UKRI Quantum Technologies for Fundamental Physics Programme [Grants ST/T005866/1 (FK) \& ST/T005858/1 (RG, SP \& TT)].
CB was supported by the UK Engineering and Physical Sciences Research Council [Grant EP/T518062/1].
The authors would also like to thank the Perimeter Institute for Theoretical Physics for hospitality.
RG also acknowledges support from the Perimeter Institute.
Research at Perimeter Institute is supported by the Government of Canada through the Department of Innovation, Science and Economic Development Canada and by the Province of Ontario through the Ministry of Research, Innovation and Science

\end{document}